\documentclass{article}
\usepackage{spconf,amsmath,graphicx,amssymb}
\usepackage{lipsum}
\usepackage[english]{babel}
\usepackage{amsthm}
\usepackage{makecell}
\usepackage{rotating}
\usepackage[hidelinks]{hyperref} 
\ninept

\theoremstyle{definition}
\newtheorem{definition}{Definition}[section]

\title{SyncNet: correlating objective for time delay estimation in audio signals}
\name{Akshay Raina and Vipul Arora\thanks{This work was supported jointly by PB/EE/2021128C grant from Prasar Bharati and IMPRINT-2C grant from SERB, Government of India.}}
\address{Department of Electrical Engineering \\
        Indian Institute of Technology Kanpur, India \\
        \texttt{\{akshayr,vipular\}@iitk.ac.in}}

\begin{document}
%
\maketitle
\begin{abstract}
This study addresses the task of performing robust and reliable time-delay estimation in signals in noisy and reverberating environments. In contrast to the popular signal processing based methods, this paper proposes to transform the input signals using a deep neural network into another pair of sequences which show high cross correlation at the actual time delay. This is achieved with the help of a novel correlation function based objective function for training the network. The proposed approach is also intrinsically interpretable as it does not lose temporal information. Experimental evaluations are performed for estimating mutual time delays for different types of audio signals such as pulse, speech and musical beats. SyncNet outperforms other classical approaches, such as GCC-PHAT, and some other learning based approaches.
\end{abstract}
\begin{keywords}
Time Delay Estimation, Convolutional Neural Networks, Correlation-based Objective, Sound Source Localization
\end{keywords}
\section{Introduction}
\label{sec:intro}

The synchronisation of signals from different sources is a typical challenge with a variety of applications, including communication \cite{ref1}, radar systems \cite{ref2}, source localization \cite{ref3,ref4}, latency estimation \cite{ref5}, self-calibration \cite{ref6} and music synchronization \cite{ref7}. A precise measurement of time delay between two signals can facilitate their synchronization. For instance, two audio devices playing the same audio need to be synchronous for a good user experience. 3D soundscape generation requires introducing precise mutual delays in multiple speakers. We focus on time delay estimation (TDE) in audio signals. We utilize it for round trip latency estimation in audio playing and recording devices. Let \(x_1(t)\) represent the reference signal to be played and \(x_2(t)\) represent the recorded signal. \(x_2(t)\) is a noisy, delayed and damped copy of \(x_1(t)\)

\vspace{-2mm}
\begin{align}
    x_2[t]=\alpha x_1[t-\tau] + w[t]
\end{align}
where \(\alpha\) is an unknown attenuation factor, 
the transmitted signal is distorted by additive noise \(w[t]\) and  \(\tau\) is the time delay between the two signals.

Many classic methods for TDE, such as cross-correlation \cite{ref8,ref9} and the generalized cross-correlation (GCC) \cite{ref10} are although insensitive to the environmental adversaries, but limited towards producing estimates only from a discrete set. Furthermore, excess noise levels in certain intervals lead to spurious peaks in GCC, which lead to poor estimates \cite{ref11}. 
Ren et al. \cite{ref13} performed onset detection by locally approximating a relevant autonomous linear state-space model (LSSM) to estimate temporal delays. 
Chen et al. \cite{refd} generalized the cross-correlation coefficient between two random signals for a multichannel setup. They proposed the multichannel cross-correlation coefficient (MCCC) method based on the spatial correlation matrix. The delay redundancy of the sensors was used for more precise estimates between the first two sensors, which was validated to be more robust towards noise or reverberation. However, MCCC is ideal for Gaussian-source signals as it is a second-order-statistics (SOS) measure of dependence among multiple random variables. 
Benesty et al. \cite{refb} overcame this limitation and established that maximising the multichannel cross-correlation coefficient (MCCC) is same as minimizing the joint entropy for Gaussian signals and showed the potential of the method having good generalizability to non-Gaussian signals like speech. 

With the recent advent of deep learning, a variety of methods for processing either raw waveforms or extracted features like spectrograms \cite{ref14, ref15} have been developed. However, using spectral domain features may harm the least count of the estimation method, rendering them unsuitable for tasks requiring high precision.
Comanducci et al. \cite{ref16} utilized a frequency-sliding GCC \cite{ref17} to make the correlations noiseless, and fed the output into an autoencoder for estimation. Wang et al. \cite{ref18} first learned a speech mask interpreted as a frequency-selective linear filter using a neural network, and then used it in conjunction with the PHAT while correlating the signals.
Salvati et al. \cite{ref19} compute multiple GCCs with distinct weighted transfer functions prior to feeding them to a convolutional neural network (CNN) for estimation.
Liu et al. \cite{refe} proposed the SCA-CRN and utilized multi-head cross-attention, layer norm, Feedforward layer and projection operations. They sought to align the far-end and near-end microphone signals using streaming cross attention before processing through a CRN network for the task of echo cancellation. 
Nauta et al. \cite{refc} used an attention based CNN to uncover causal linkages in observational time series data and build a causal graph structure. 
Berg et al. \cite{ref20} suggested filtering signals with neural networks prior to estimating time delay using the GCC-PHAT. They classify into only a set of possible time delays using Cross Entropy loss function, which is undesirable for a number of real-world applications as the possible delays must be defined prior to network training. However, the promising results motivate to utilise neural networks, although the methodology suffers with undesirable least-count of the estimates, i.e., it cannot be employed in cases requiring high precision. 

This paper presents a novel method for high-precision time-delay estimation by mapping the input signals into a latent space using a deep neural network with its objective function derived from the cross-correlation of the transformed signals. The correlation of the resulting sequences in latent space peaks at the correct time delay. This makes the methodology intrinsically interpretable as it builds on signal processing methods.


\section{Proposed Methodology}
\label{sec:meth}

\subsection{Latent Space Transformation}
SyncNet uses a 1-dimensional convolutional neural network to transform the input signals to a latent space where they could be correlated to estimate time delay. Convolutional Neural Networks are supposedly shift-equivariant, as defined below. This allows SyncNet to capture local temporal information of the input signal, thereby generating representations that could be used for estimating time delay.

\begin{definition}[Shift-Equivariance]
A function $f:\mathbb{R}^{N}\rightarrow\mathbb{R}^N$ is said to be shift-equivariant if $f(x[t+\tau])[t']=f(x[t])[t'+\tau]$ for any $\tau\in\mathbb{Z}$.
\end{definition}
The reference and the noisy-delayed signals are passed through the same 1-D neural network \(f_\theta\) along the channels dimension. The transformed signals from the network are then correlated to give \(\hat{R}(\tau)\). The network learns to map the input signals with reverberation and noise to other signals such that the cross-correlation sequence peaks at the actual delay between the two input signals. To keep the size of the output signal same as the input, SyncNet does not use any pooling or non-unity strides along the time axis.

\begin{figure}[t]
  \centering
  \includegraphics[width=\columnwidth, height=2.3cm]{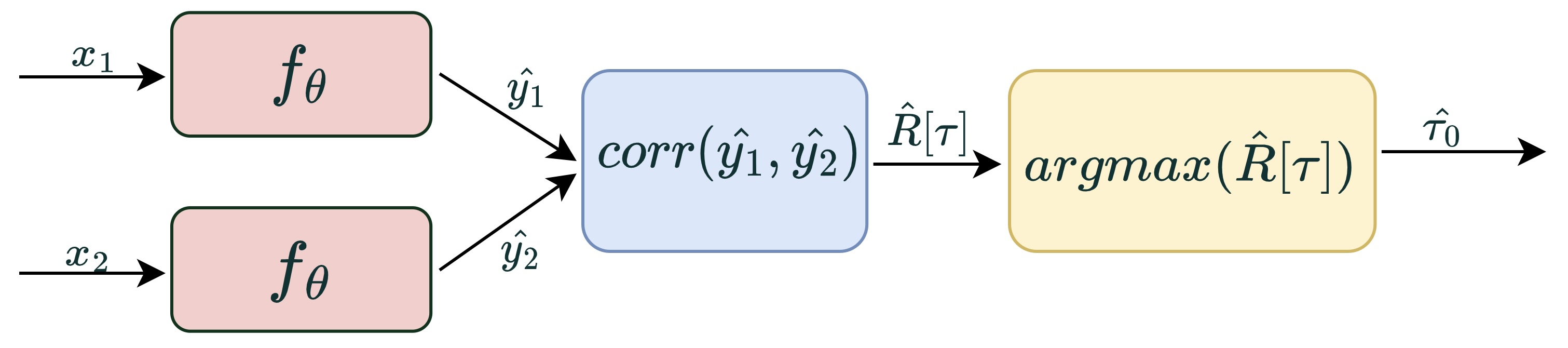}
  \caption{Block diagram of the overall architecture of SyncNet}
  \label{fig1}
\end{figure}

In order to reduce the number of samples to be passed through the objective function and considering the scope of affording some imprecision in estimates, SyncNet pools the correlation sequence of the transformed sequences. This can be better analysed by an overview of the proposed methodology in Figure~\ref{fig1}.
\vspace{-0.8mm}
\begin{align}
	\label{equation5}
	\hat{R}[\tau] = corr(\hat{y_1}[t], \hat{y_2}[t])
	\;\text{;}\;
	\hat{y_i}[t] = f_\theta(x_i[t])\; \forall\;i \subset {\{1,2\}}
\end{align}

For time-delay estimation, one may desire to transform the input signals to their noiseless, less-reverberation variants. However, for several real-world use cases, the definition of background noise is subjective to the database and task. As discussed in detail in section 3.2, the learned representations of input signals by SyncNet can be interpreted as containing the onsets of the event in the audio. This makes SyncNet less black-box as obtaining the onsets of events in two audio files can be trivially related to the time delay of the events.

\vspace{-1.2mm}

\subsection{Objective Function}
\label{objectivefunction}
The cross-correlation function of two signals \(x_1[t]\) and \(x_2[t]\), where \(E[.]\) denotes the expectation is defined as-
\begin{equation}
\label{equation3}
R_{x_1x_2}[\tau] \overset{def}{=} E[x_1[t]x_2[t-\tau]]
\end{equation}
The argument \(\tau=\tau_0\), which maximizes \(R_{x_1x_2}[\tau]\) corresponds to the estimate of temporal delay in two signals. Several studies \cite{ref9,ref10} have shown that using the cross-correlation function directly or upon some filtered signal variant results in acceptable performance as a time delay estimator. Furthermore, Berg et al. \cite{ref20} showed that injecting domain-specific knowledge into a deep learning system significantly improves the performance of the estimator. With this motivation, we utilized the cross-correlation sequence of the transformed sequences for training the neural network.

Let \(\hat{y_1}[t]\) and \(\hat{y_2}[t]\) be the predicted sequences for the input noisy-delayed and reference signals, \(x_1[t]\) and \(x_2[t]\) respectively, by the network. The cross-correlation sequence of both transformed sequences can be denoted by \(\hat{R}_{y_1y_2}[\tau]\). 
The network should be trained to transform the input sequences such that \(\hat{R}_{y_1y_2}[\tau]\) peaks at the actual delay in time. To achieve this, the loss function is formulated as a regression loss to match the cross-correlation sequence to a Gaussian sequence, which peaks at the actual time delay. In the case of periodic signals, their cross-correlation function will also be periodic, leading us to use a sequence of Gaussians. This can be devised as-

\begin{equation}
    \label{equation4}
    R(\tau)=\sum_{n=0}^{g-1}{\frac{1}{\sigma_n\sqrt{2\pi}}\exp{-\frac{{(\tau-\mu_n)}^2}{2\sigma_n^2}}}
\end{equation}
where \(\mu_n = T_0 + nT\) is the \(n^{th}\) onset, and \(\mu_0 = T_0\) is the actual delay in time and $T$ is the time period of the reference signal. Note that for aperiodic signals, where the cross-corrleation is also generally aperiodic, the second term \(nT\) can be ignored. For this context and simplicity, we can set \(\sigma_n = \sigma\) \(\forall\) \(n \in \{0,1,2,...(g-1)\}\). It is desirable for TDE that the correlation between two signals has clear and distinguishable peak. 

Let \(\mathcal{L}\) be a distance metric between \(R[\tau\)] and \(\hat{R}_{y_1y_2}[\tau]\). 
We define \(\mathcal{L}\) as
\begin{align}
    \label{equation6}
    \mathcal{L}(.) \overset{def}{=} l_1\mathcal{L}_1(.) + l_2\mathcal{L}_2(.)+ l_3\mathcal{L}_3(.)
\end{align}
Here, \(\mathcal{L}_1=\sum_{i=1}^{N^{'}}{{(R_i-\hat{R}_i)}^2}\) is a simple MSE function, \(\mathcal{L}_2=\sqrt{\sum_{i=1}^{N^{'}}{{(\log{R_i^{\wp}}-\log{\hat{R}_i^{\wp}})}^2}}\) is the root-mean-log error function, \(\mathcal{L}_3=\sum_{i=1}^{N^{'}}{R_i^{\wp}(\log{R_i^{\wp}}-\hat{R}_i^{\wp})}\) is the KL-Divergence loss and \(l_i \forall i \in \{1,2,3\}\) are the corresponding weights associated to each term. The obtained cross-correlation sequence (\(\hat{R}(\tau)\)) is pooled down by $\wp$ samples to get \(\hat{R}^{\wp}(\tau)\) such that the most activated sample for every \(\wp\) samples in \(\hat{R}(\tau)\) is chosen as \(i^{th}\) sample in \(\hat{R}^{\wp}(\tau)\). This is done primarily to reduce the number of samples in \(\hat{R}(\tau)\) lacking the required peak, thereby assisting with the peak/no-peak imbalance problem. Also, the choice of $\wp$ affects the precision of the time alignment algorithm as the least count of the algorithm will be $\wp T_s$, where $T_s$ is the sampling time of the input.

Note that, \(\mathcal{L}_1\) is applied onto the correlation sequences, thus helps with matching its shape with target, whereas, \(\mathcal{L}_2\) and \(\mathcal{L}_3\) are operating over the pooled sequences. 
We employed the MSE loss ($\mathcal{L}_1(.)$), simply because of the regression-task. However, to aid in learning of clear peaks in correlation-sequence we also utilized the RMSLE ($\mathcal{L}_2(.)$), which is a well-known regression-loss measure. The reason being two-fold; first, unlike $\mathcal{L}_1(.)$, it penalizes a prediction relative to the corresponding ground-truth. Second, it penalizes underestimation of the actual value more harshly than it does for the overestimation. To have its influence higher in the overall loss, $l_1$ is set lower than $l_2$. The KL-divergence ($\mathcal{L}_3(.)$) was utilized with the understanding that mininimizing $\mathcal{L}_3(.)$ would help with making the predicted correlation-sequence equal to a Gaussian density function, which is desirable. It is trivial to establish that $D_{KL}(P||Q)=0~\Leftrightarrow~P(x)=Q(x)$ as measures.

Thus, the estimated parameters, where \(\ast\) represents the cross-correlation operation are-
\begin{align}
    \label{equation7}
    \hat{\theta} = \underset{\theta \in \Theta}{min\ }\mathcal{L}(R(\tau),\hat{y}_1[t]\ast\hat{y}_2[t])
\end{align}

The use of cross-correlation function to form a training objective has not been explored before, as far as we know. The closest we know is the concept of maximizing cross-correlation for static features, not time sequences, in the case of deep canonical correlation analysis of multi-modal data \cite{ref21}.

Clearly, the obtained sequence \(\hat{R}(\tau)\) is expected to have far lesser samples with a peak than without one. This peak/no-peak imbalance issue can be tackled by weighing the loss function at particular indices. Therefore, we up-weighted \(\mathcal{L}\) at \(\tau=\tau_0 + nT\) by a scalar \(u\) and down-weighted for other indices by another scalar \(d\). The values of these constants are algebraically computed as-
\vspace{-1.5mm}
\begin{align}
	\label{equation5}
	d \leftarrow 1 - \frac{g+1}{\lambda}
	\;\;\text{and}\;\; 
	u \leftarrow d + \frac{N^{'}}{\lambda}
\end{align}

\begin{table}[t]
  \caption{Network architecture studied. Here \(c\) represents the number of Conv1D kernels and \(k\) is the size of these kernels. Part refers to a sequential module of one or more layers}
  \label{tab1}
  \centering
  \begin{tabular}{c|c|c}
    \textbf{Network Part} & \textbf{Number of parts} & \textbf{Hyperparameters}\\
    \hline
      Conv1D &  {} &  \(c \in \{16,32,64,128,256\}\)\\
    BatchNorm1D  & 10 &  \(k = 61\)\\
    ReLU &  {} &  {}\\
    \hline
    {} &  {} &  {}\\
    Conv1D & 1 &  \(c=2\)\\
    {} & {} &  {}\\
  \end{tabular}
\end{table}

\begin{table}[t]
\centering
\caption{Test of generalizability for SyncNet. First row represents the performance (MSE in $sec^2$) when models were trained on Synthetic and tested on real recordings, while results with training on Librispeech and testing on MTic is tabulated in second row}
\label{tab2}
\begin{tabular}{c|c|c|c|c}
                                        & \multicolumn{4}{c}{\textbf{Method}}             \\ 
\hline
{} &    \textbf{\textbf{Train-Test}}         & \textbf{AfC} & \textbf{MSfC} & \textbf{SyncNet}  \\ 
\cline{2-5}
                                \rotcell{\textbf{Split}}        & Synthetic-Real        & 0.107   & 0.088          & \textbf{0.060}      \\
                                     & Librispeech-MTic & 0.073   & 0.065          & \textbf{0.047} 
\end{tabular}
\end{table}

Here, \(\lambda=10^a\) for the smallest number \(a\) such that \(\lfloor \frac{N^{'}}{10^a} \rfloor = 0\) and \(N^{'}\) is the number of samples in \(\hat{R}_{y_1y_2}(\tau)\). 

\section{Evaluation}
\subsection{Dataset Used}
We evaluate the proposed method on a variety of datasets - including both real recordings as well as synthetically generated audio.  
We prepared a new dataset called MTic. A reference audio signal of periodic tics with a time period of $1$s is played on a phone speaker. It is recorded by the microphone of the same phone.
Tics are chosen for experiments because they are localized in time and are musically relevant (as in metronome). The recordings are done with a number of phones and in various acoustic background conditions.
The time delay in the recorded signal is $<0.9$s.
There are 170 audio recordings sampled at \(16kHz\), each of a duration of almost \(10\) seconds. They all correspond to the same reference audio of $10$s duration.
Furthermore, to increase the sample size, we synthetically generate 450 audio files from the same reference signal with delays chosen uniformly between \((0s,0.9s]\) and noise injections with signal-to-noise ratio varying in \([0dB,30dB]\). 

To validate the robustness of this study over real-world signals such as speech and music, we experiment with speech files from the LibriSpeech dataset \cite{ref22}, which contains speech recordings from read audiobooks in English, sampled at 16kHz; speech is non-periodic. We sampled 10000 speech audio files from the database, and 80:20 split on top of these files was used for training and validation. For each epoch, a noisy and delayed signal was synthesized with an SNR in \([0dB,30dB]\) and delayed by an amount in \((0.2s,3s]\) from each recording in the training set. Therefore, the network was trained on different reference-delayed pairs, sampled from the same process.
We also experiment with the accompaniment music from MIREX 2012 dataset \cite{ref23}, which we call MBeats dataset. 
The MTic dataset as well as the codes to reproduce the results in this paper are available at
\texttt{https://github.com/madhavlab/2022\_syncnet}.


\begin{figure}[t]
  \centering
  \includegraphics[width=\columnwidth,height=7.6cm]{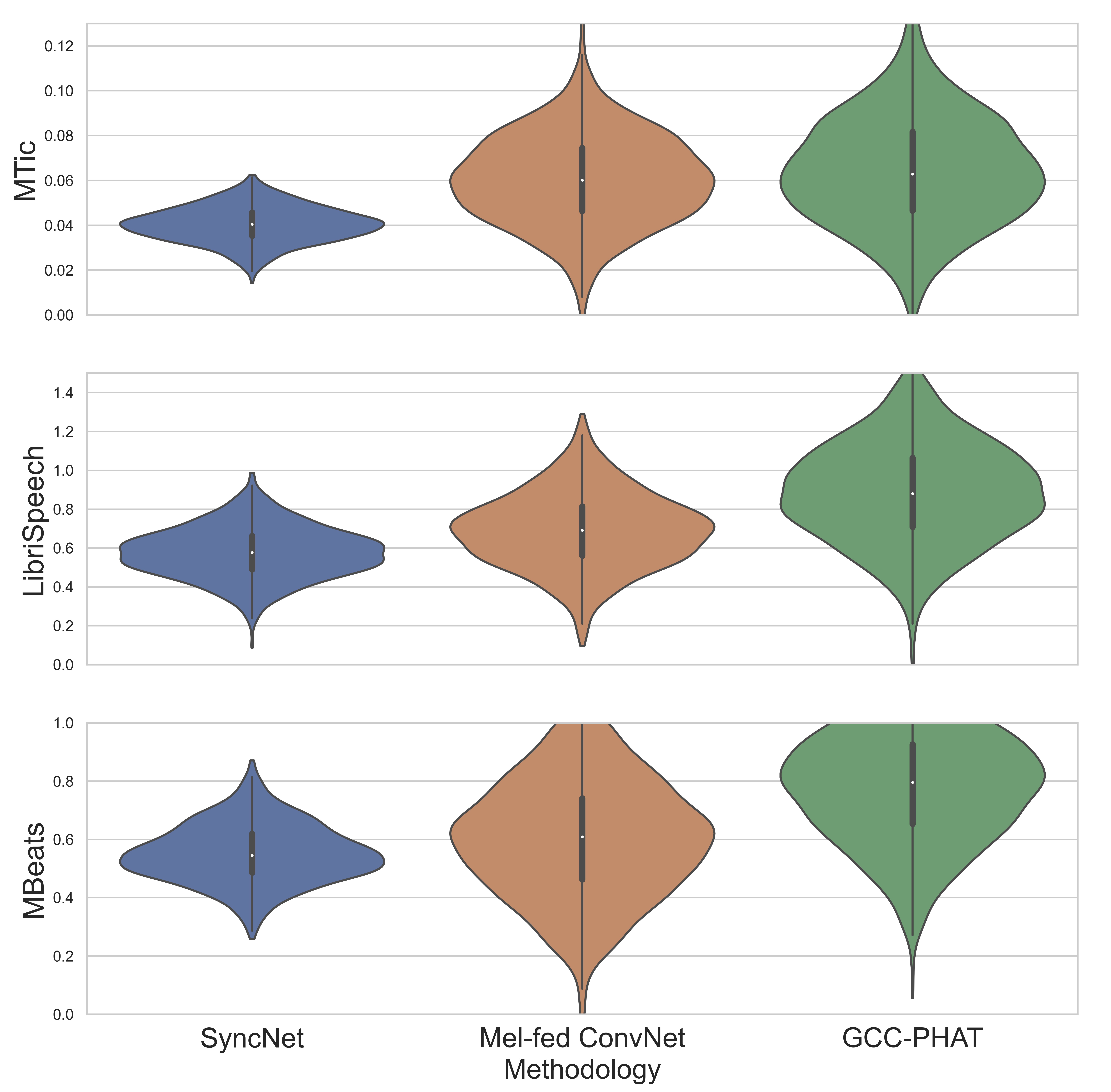}
  \caption{Violin plots for absolute error for GCC-PHAT, MfCN and SyncNet on all 3 datasets. Vertically lower distribution indicates better performance translats to.} 
  \label{fig3}
\end{figure}

\subsection{Experiments}
\label{exp}

\begin{figure}[t]
  \centering
  \includegraphics[width=\columnwidth,height=4.4cm]{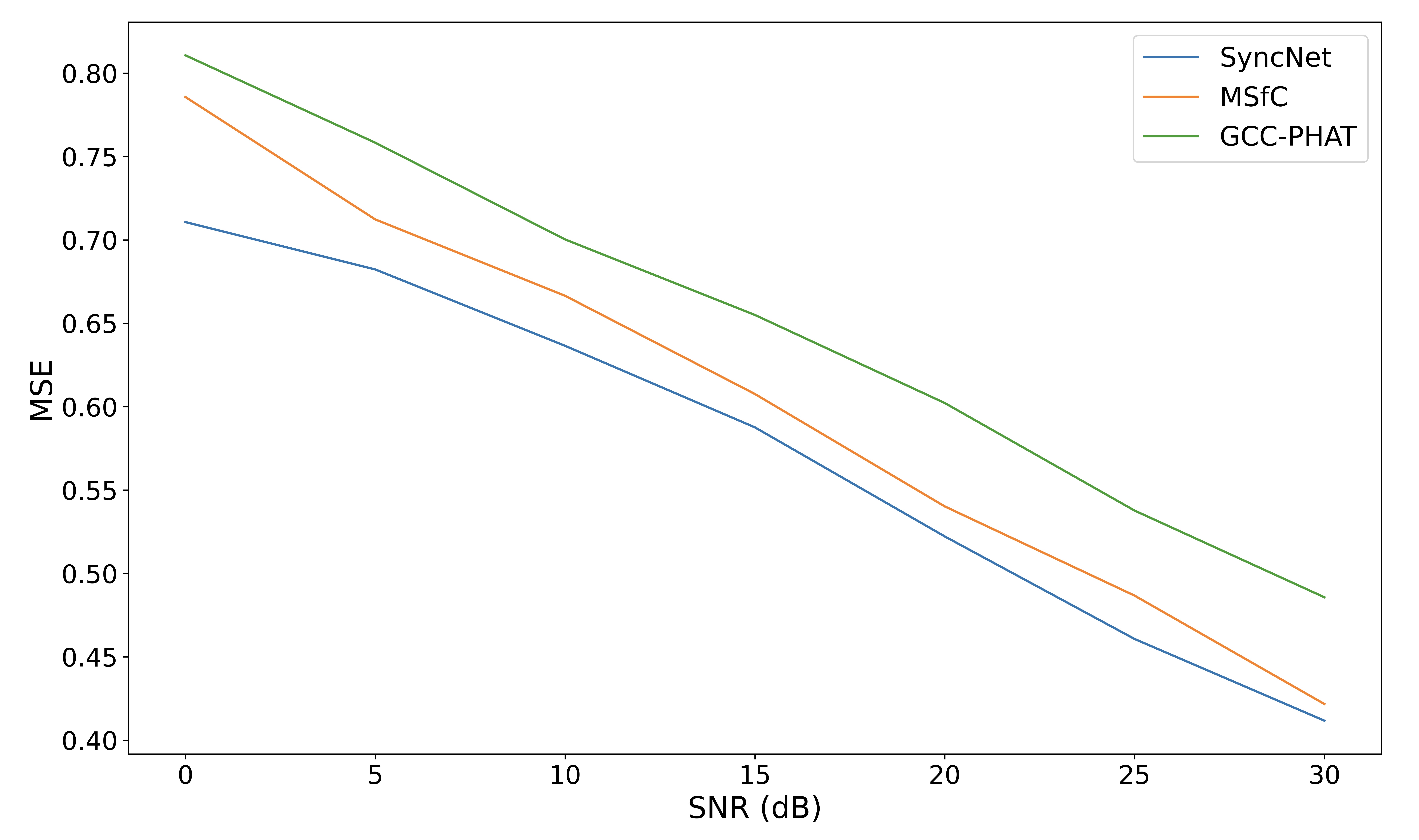}
  \caption{The MSE for the Time Delay estimation performance on LibriSpeech at different SNR levels.} 
  \label{fig2}
\end{figure}

We have conducted several experiments to validate the robustness of SyncNet. Both reference and signal-of-interest need to be transformed into embeddings, trained for the task at hand. For this, both signals are passed along the channel dimension of the same 1-D network.
There are no pooling layers in the architecture. Batch-Normalization layers are used, and layers are activated using the ReLU function as shown in Table~\ref{tab1}. 
 The pool size (\(\wp\)) has been set to 60. The network was trained for 50 epochs in all experiments using a linear learning rate scheduler and Adam optimizer. 
The mean squared error between the estimated value of delay and the actual delay value is naturally chosen as the evaluation metric. 
For the baseline, we train a deep 1-D audio fed-ConvNet (AfC), a Mel-Spectrogram fed ConvNet (MSfC) with similar layers but linear layers in the end with sigmoid activation and Mean Squared Logarithmic Error Loss function. Although there have been numerous attempts to solve TDE using Signal Processing/ Deep Learning [12-19] but mostly on simulated sensor data, which is not available publicly. Our study sets a benchmark as it discusses performance on well-known databases. AfC and MSfC are naive baselines using conventional/vanilla deep learning approaches. They may be seen as ablated models to study the importance of the proposed correlation-based objective function and hybrid loss.
Table~\ref{tab3} shows the mean squared error (MSE) for all the methods on the three datasets -- MTic, Librispeech, and MBeats. We can see that SyncNet consistently outperforms all the baseline methods over all the datasets in terms of average performance. Figure~\ref{fig2} validates the robustness of the methodology towards variation in SNR. SyncNet is relatively much more stable than the baseline methods.


\begin{table}[t]
  \caption{Mean squared error in time delay estimation, obtained for all the methods on the three datasets (in $sec^{2}$)}
  \label{tab3}
  \centering
  \begin{tabular}{c|c|c|c}
    \textbf{Method} & \textbf{MTic} & \textbf{LibriSpeech}& \textbf{MBeats} \\
   \hline
      Cross-Correlation &  0.091&  1.064 & 1.023\\ 
      GCC-PHAT &  0.067 & 0.885& 0.817\\
      AfC &  0.067 & 0.810&  0.731\\ 
      MSfC &  0.061&  0.697 & 0.607\\ 
      SyncNet &  \textbf{0.041} &  \textbf{0.580}& \textbf{0.531}\\ 
  \end{tabular}
\end{table}

However, to dive deeper into the performance of the three top performing methods, we draw the violin plots to see the distribution of errors across all test samples.
Figure~\ref{fig3} shows these violin plots. It can be seen that the estimation errors for the proposed SyncNet method are distributed more towards the lower end as compared to other methods, supporting the improvements brought in it.
It is important to note that the proposed method is not limited by the length of the signals, and performs consistently with signals of variable lengths as well as sampling rates, maintaining its higher precision.

Since we experiment on a variety of databases with both synthetic and real recordings, it naturally requires a generalizability test for SyncNet. For this, we first trained the three deep neural networks on synthetic recordings and tested them on real recordings for the MTic database. Second, for cross-dataset generalization, we trained the networks on Librispeech and tested them on recordings from MTic database. The obtained results as tabulated in Table~\ref{tab2} clearly indicate that for most real-world use cases, SyncNet is capable to generalize while producing precise estimates. Lastly, as discussed in Section~\ref{objectivefunction}, the terms in the loss function used have distinct impact on the overall training of the network. To validate the importance of the hybrid loss, we performed an ablation study on the effect of the three losses on the performance of SyncNet on all three databases. The results are reported in Table~\ref{tab4}.

\begin{table}[t]
  \caption{Ablation study on effect of $L_i$ on the performance (MSE in $sec^2$) of SyncNet}
  \label{tab4}
  \centering
  \begin{tabular}{c|c|c|c}
    \textbf{Loss function} & \textbf{MTic} & \textbf{Librispeech} & \textbf{MBeats}\\
   \hline
      $\mathcal{L}_1(.)$ & 0.070 & 0.983 & 0.904\\ 
      $\mathcal{L}_2(.)$ & 0.051 & 0.713 & 0.660\\
    $\mathcal{L}_3(.)$ & 0.058 & 0.804 & 0.701\\ 
      $l_1\mathcal{L}_1(.) + l_2\mathcal{L}_2(.)+ l_3\mathcal{L}_3(.)$ &  \textbf{0.041} &  \textbf{0.580}& \textbf{0.531}\\
  \end{tabular}
\end{table}

\begin{figure}[t]
  \centering
  \includegraphics[width=\columnwidth,height=4.8cm]{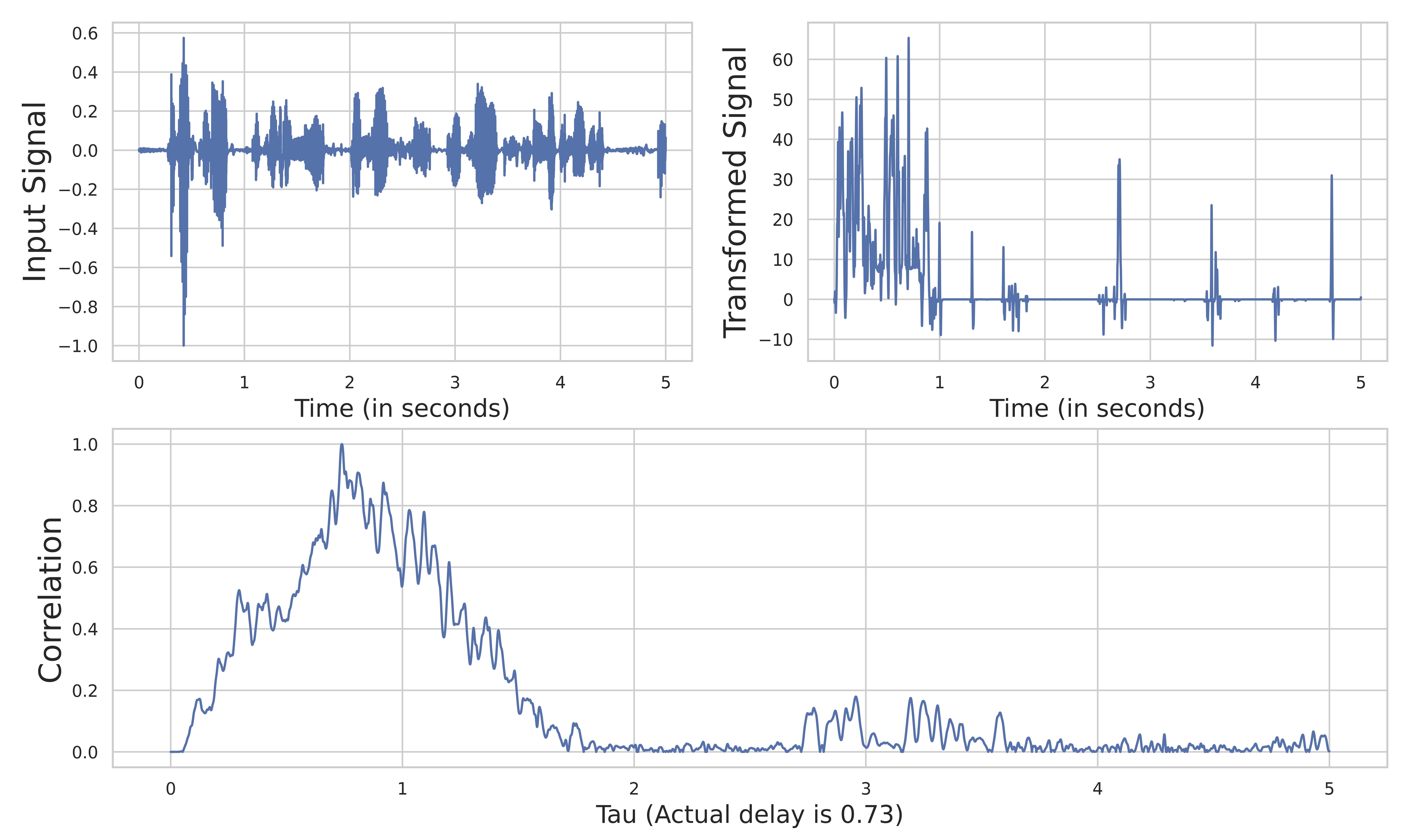}
  \caption{Input delayed-noisy signal, its transformed variant and pooled correlation sequence estimated. The correlation sequence peaks correctly at the actual time delay, i.e., 0.73s.} 
  \label{fig4}
\end{figure}

The Signal Processing based methods \cite{ref9,ref10} are considered more explainable, and thus have a higher potential for fine tuning and debugging. Deep Learning based techniques however, mostly lack explainability. Nonetheless, SyncNet transforms a pair of input sequences into a pair of sequences, such that a simple signal processing metric, such as cross-correlation may be used to estimate their mutual delay. This makes the network explainable for its estimates, as the causation of predictions of the network are well understood. As an example, Figure~\ref{fig4} shows a reference signal from LibriSpeech database as input to SyncNet, the output of SyncNet and its cross-correlation with that of the corresponding signal-of-interest. A clear peak can be noticed at the actual delay in the correlation plot.

\section{Conclusion}
\label{conclusion}
We introduce a novel correlation function based objective function for deep neural network. It utilizes time shift equivariance for transforming the input signals into useful representations. The embeddings with the captured temporal information about the input signal can thus be used for finding peaks in the correlation sequence. The experiments validate the robustness of SyncNet, which outperforms many existing approaches. We are working towards applying this method to problems such as round trip latency estimation for audio playback and recording on mobile devices. This synchronized audio can be used by an online service provider to learn the distortion as well as time delay characteristics of these devices in the wild.

\bibliographystyle{IEEEbib}
\bibliography{refs}

\end{document}